\pdfoutput=1
\documentclass[graybox]{svmult}

\usepackage{mathptmx}       % selects Times Roman as basic font
\usepackage{helvet}         % selects Helvetica as sans-serif font
\usepackage{courier}        % selects Courier as typewriter font
\usepackage{type1cm}        % activate if the above 3 fonts are
\usepackage{url}
\usepackage{makeidx}         % allows index generation
\usepackage{graphicx}        % standard LaTeX graphics tool
\usepackage{multicol}        % used for the two-column index
\usepackage[bottom]{footmisc}% places footnotes at page bottom

\usepackage{amssymb}
\makeindex             % used for the subject index
\begin{document}

\title*{Fluctuations in pedestrian evacuation times: Going one step beyond the exit capacity paradigm for bottlenecks}
\titlerunning{Fluctuations in pedestrian evacuation times}
\author{Alexandre NICOLAS}
\institute{LPTMS, CNRS, Univ. Paris-Sud, Universit\'e Paris-Saclay, 91405 Orsay,
France \email{alexandre.nicolas@polytechnique.edu}}

\maketitle

\abstract{
For safety reasons, it is important that the design of buildings and public facilities comply with the guidelines compiled in building codes.
The latter are often premised on the concept of exit capacity, \emph{i.e.}, the mean pedestrian flow rate through a bottleneck (at congestion). Here, 
we argue that one should duly take into account the evacuation time fluctuations when devising these guidelines. This is particularly true when the narrowing is
abrupt and the crowd may behave competitively. We suggest a simple way to assess 
the extent of (part of) these fluctuations on the basis of the statistics of time gaps between successive escapes through the considered
bottleneck, which in practice could be garnered by analysing recordings of future real evacuations or, perhaps, realistic drills (in the limits of what is ethically possible). We briefly present a test of the proposed strategy using a cellular automaton model and confirm its validity under some conditions, but also
disclose some of its limitations. In particular, it may severely underestimate fluctuations in the presence of strong correlations in the pedestrians' behaviours
(while still performing better than only the mean capacity).
}

\begin{center}
(Submitted to the \emph{Proceedings of the 2017 Traffic and Granular Flow conference}, to be published by \emph{Springer})
\end{center}

\section{Introduction}

In emergency situations (whether it be a fire outbreak, an earthquake,
a terrorist attack, etc.), buildings and public facilities usually
need to be evacuated quickly and safely. Building architectures should
be designed accordingly. In particular, corridors and doorways on
the egress pathway must be wide enough to allow the occupants to egress
without excessive delay and to reduce the risk of clogging at the
bottlenecks. How can this plain condition translate into a practical guideline? Building codes answer this question either by prescribing
minimal requirements in terms of bottleneck widths or by setting the minimal standards
(performances) that should be met with respect to the evacuation time. 

To take a few examples, building codes in the United States (for instance
in Florida \cite{florida2010florida}) prescribe a minimal clear
width 
$w=\max\left\{ 813,\,5.1\,N\right\}\, \mathrm{mm}$ for doorways,
where $N$ is the occupant load. Note that $w$ is lowered to $\max\left\{ 813,\,3.8\,N\right\} \,\mathrm{mm}$
in suitably equipped low-risk buildings. In France, emergency exit doorways and corridors
in public buildings must typically be larger than $\left(1+\left\lceil \frac{N}{100}\right\rceil \right)\times0.6\,\mathrm{m}$
if $200<N\leqslant500$, where $\left\lceil x\right\rceil \in\mathbb{N}$
is the ceiling of $x$, or $\left\lceil \frac{N}{100}\right\rceil \times0.6\,\mathrm{m}$
if $N>500$ \cite{legifrance2017}. French railway stations must be designed in such a way
that they can be evacuated in less than 10 minutes. To evaluate the
evacuation time, a walking velocity $v=1.4\:\mathrm{m\cdot s^{-1}}$
is assumed, along with a specific capacity $J_{s}=1.67\,\mathrm{m^{-1}\cdot s^{-1}}$,
\emph{i.e.}, a maximal flow rate of 1.67 people per second per meter of corridor
width. These figures are reduced to $v=1.0\:\mathrm{m\cdot s^{-1}}$
and $J_{s}=1.0\,\mathrm{m^{-1}\cdot s^{-1}}$ if it is a regional
or national station \cite{legifrance2017}.
Finally, in the United Kingdom, the maximal evacuation time for
sports stadiums is set to a value between 2.5 and 8 minutes, depending
on the risks, and a specific capacity $J_{s}=1.37\,\mathrm{m^{-1}\cdot s^{-1}}$
is assumed for level walkways \cite{football2008guide}.

These guidelines are thus premised on the concept of exit capacity,
or in other words mean flow rate in saturated conditions. Not surprisingly, the values
indicated in the above building codes are slightly lower, but comparable
to the values measured in controlled experiments where participants
were asked to walk through a bottleneck in normal conditions (for instance, $J_{s}=1.85\,\mathrm{m^{-1}\cdot s^{-1}}$
was reported in \cite{kretz2006experimental}. On the other hand, they are considerably lower than the values
measured in controlled competitive conditions through narrow doors: In \cite{Pastor2015experimental},
Zuriguel \emph{et al.} measured $J_{s}\approx3.7\,\mathrm{m^{-1}\cdot s^{-1}}$.
Varying the participants' (prescribed) eagerness to escape,
my colleagues and I observed the whole range of values $1.4\leqslant J_{s}\leqslant3.3$ (in $\mathrm{m^{-1}\cdot s^{-1}}$) \cite{nicolas2017pedestrian}.
The underestimates provided in building codes can be interpreted as safety margins, intended
to absorb unforeseen delay. But is this enough to ensure that the
evacuation time will always be within the chosen bounds?

Here, we show that the presence of significant fluctuations,
of diverse origins, undermines this reasoning based on mean values
(Section~\ref{sec:fluctuations}). We then propose a simple method
to assess these fluctuations in Section~\ref{sec:micro-macro} and test it in Section~\ref{sec:validity}.

\section{Beyond the mean exit capacity: Fluctuations\label{sec:fluctuations}}

\subsection{Importance of fluctuations}

Since the dynamics of evacuations depend on many details that are
out of control, significant fluctuations should be foreseen. They
undermine any reasoning based exclusively on mean values. Indeed,
in the presence of strong fluctuations, knowing that evacuations are
quick enough \emph{on average }does not tell you how often they will
be excessively lengthy. Even if we focus only on the delays expected
at bottlenecks (doorways, relatively narrow corridors, etc.), experiments
involving mice that were forced to flee through a narrow orifice showed
that realisations conducted in virtually identical conditions exhibit
a significant dispersion, with standard deviations (std) that typically
amount to 6\% to 14\% of the total evacuation times of about 90 mice,
for a variety of settings \cite{lin2016experimental,lin2017experimental}
(the figure is closer to the upper bound when there are many
realisations). Similar figures are reported for the entrance of 85 sheep into a barn through a narrow gate,
with an std-over-mean ratio of around 15\% for the total time \cite{garcimartin2015flow},
and the importance of the dispersion of the evacuation times was underscored
by the authors. Considering a crowd of about 90 participants walking
through a narrow (69-cm-wide) door in a controlled experiment, the
ratio is found to be around 7\%, with moderately competitive participants
as well as with highly competitive ones  \cite{garcimartin2016flow}. (Note that the given std-over-mean
ratios are the results of my own calculations using the data of the
cited papers.) Importantly, even though intermittence in the dynamics is
favoured by the presence of abrupt narrowings, the fluctuations
 do not vanish when
the constriction gets wider and they may be very considerable if the crowd
behaves frantically. A striking example is the huge clog that occurred
in the running of the bulls during the 2013 San Ferm\'in in Pamplona,
when the crowd running in front of the bulls pushed so hard on the
human clog formed at the entrance of the arena that, despite its being
a few meters wide, not even a handful of people could enter per second. This very large deviation from the mean flow rate is a sign
of anomalously broad statistics.

As a consequence, it is of paramount interest to have an idea of the
distribution of evacuation times, beyond its mean value. This
distribution corresponds to an ensemble of realisations, for a fixed
number of occupants $N$ and a fixed geometry. Unfortunately, it is
unrealistic to expect the collection of such statistics for any $N$
and any geometry. In the following, we will propose a strategy to
bypass this need. To this end, we will first disentangle the origins
of the fluctuations.

\begin{figure}
\noindent \begin{centering}
\includegraphics[width=\textwidth]{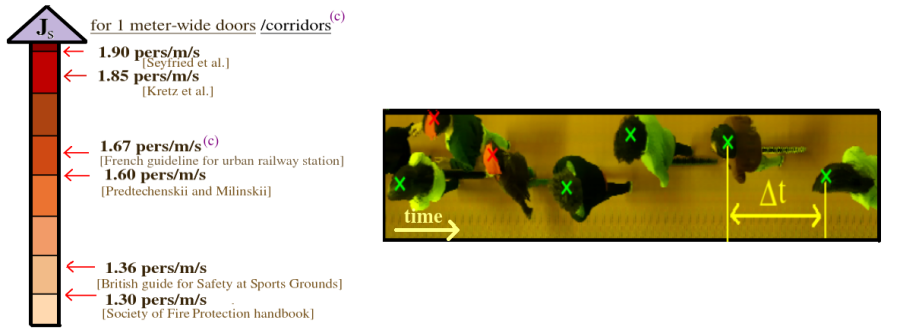}
\par\end{centering}

\caption{\label{fig:figure1}(\emph{Left} Some guidelines or measurements for the specific capacity of doors and corridors (\emph{i.e.}, the flow
rate per metre of cross section). (\emph{Right}) Time line of a controlled evacuation \cite{nicolas2017pedestrian} obtained by stitching
pixel lines corresponding to the doorway in successive frames; the definition of the time gap $\Delta t$ is illustrated. }
\end{figure}

\subsection{\label{sec:origins}Origins of the fluctuations}

\textbf{\emph{Statistical fluctuations. }} As any complex system, pedestrian crowds feature an amount of disorder:
The pedestrian's morphologies and their preferential velocities are
different, their initial positions at the beginning of the evacuation
are more or less random, etc. For these reasons, successive realisations
(even conducted with identical crowds and in seemingly identical conditions)
will not display exactly the same dynamics. The observed differences
are statistical fluctuations due to the stochasticity of uncontrolled
parameters and are also present in the flow of purely mechanical systems,
such as grains discharging from a silo \cite{Lozano2012flow}.

\textbf{\emph{Extrinsic variations. }} Besides these statistical fluctuations, extrinsic variations are also
expected in real evacuations. In particular, the composition of the
crowd that might need to evacuate the building is not the same from
day to day, nor will their eagerness or competitiveness to egress
be the same. Turning up the last parameter, for instance, was shown
to produce more intermittent (bursty) evacuation dynamics \cite{Pastor2015experimental,nicolas2017pedestrian}.

\section{A practical method to assess statistical fluctuations\label{sec:micro-macro}}

Having disentangled the origins of the fluctuations, we now propose
a practical way to predict the statistical fluctuations of the global
evacuation time.

\subsection{Distribution of time gaps between successive egresses}

Consider a given escape zone. Let $t_{i}$ be the time of the $i$-th
escape (out of $N$) in a realisation of the evacuation and let $\Delta t_{i}$
be the time gap $t_{i}-t_{i-1}$, where we have set $t_{0}=0$, as illustrated in Fig.~\ref{fig:figure1}. The
premise of our approach is that it is possible to collect enough statistics
about these time gaps to get a decent approximation of their distribution
$p(\Delta t)$ for a typical crowd composition. This requires the
observation of a reasonable number of real evacuations, which may
be achieved in the near future, owing to the increased monitoring
of public facilities. Alternatively, one may choose to perform evacuation
drills in conditions as realistic as is ethically possible.

\subsection{Micro-macro relation}

To proceed, we remark that the total evacuation time of the $N$ occupants
(\emph{i.e.}, here, the delay at the bottleneck) reads 
\[
T_{\mathrm{esc}}(N)=\sum_{i=1}^{N}\Delta t_{i}.
\]
It can thus be regarded as a sum of $N$ random variables drawn from the
distribution $p(\Delta t)$. If we overlook possible correlations
between successive time gaps, the distribution of $T_{\mathrm{esc}}(N)$
is given by the following \emph{micro-macro} \emph{relation}, based
on a convolution product ($*$), 
\begin{equation}
P_{N}\left(T_{\mathrm{esc}}\right)=p^{*N}\left(T_{\mathrm{esc}}\right).\label{eq:random_sum-1}
\end{equation}

\begin{svgraybox}
In particular, provided that the `microscopic' distribution $p(\Delta t)$
has a finite mean $\overline{\Delta t}$ and a finite standard deviation
$\sigma,$ the central limit theorem implies that, in the limit of
large attendance $N\gg1$, $P_{N}\left(T_{\mathrm{esc}}\right)$ follows
a normal law of mean $N\overline{\Delta t}$ and of variance $N\sigma^{2}$.
Thus, we have managed to assess the global distribution $P_{N}\left(T_{\mathrm{esc}}\right)$,
whose direct assessment would require a hopelessly large amount of
data because of its dependence on $N$, on the basis of the more accessible
$p(\Delta t)$ \cite{nicolas2016statistical}.
\end{svgraybox}

\begin{figure}
\noindent \begin{centering}
\includegraphics[width=1\textwidth]{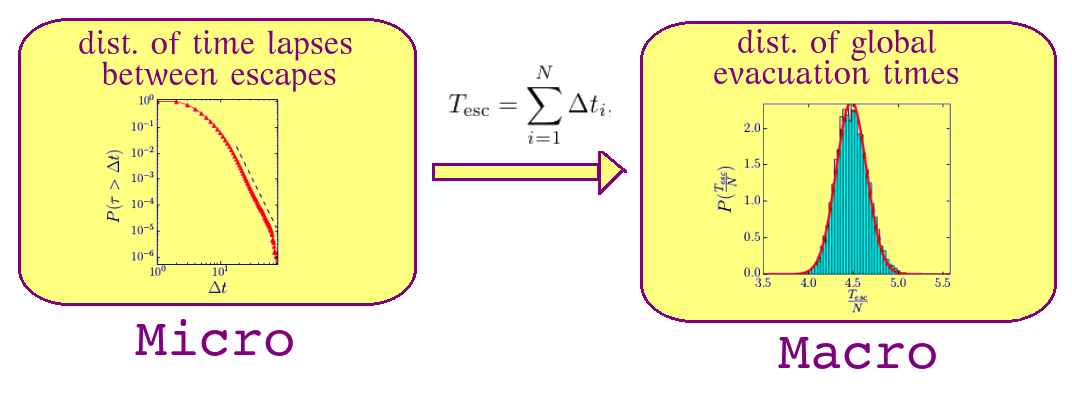}
\par\end{centering}

\caption{ \label{fig:micro-macro} Illustration of the \emph{micro-macro relation}, which takes as input the
`microscopic' statistics of time gaps $\Delta t$ at the door and
predicts the global distribution of evacuation times $T_{\mathrm{esc}}$
for the $N$ occupants.}
\end{figure}

\subsection{Caveats and possible issues}

Already at this stage, we ought to mention some caveats with respect
to the \emph{micro-macro relation}. First, one may think that $p(\Delta t)$
actually depends on $N$. But we believe that, granted that $N$ is
large enough (typically $N>100$ for a narrow door \cite{garcimartin2016flow}),
this dependency can be overlooked (unless huge pressure builds up
at the door, in which case pressure itself will cause a tragedy, regardless
of the delay at the door). Indeed, in the controlled experiments we
performed (see \cite{nicolas2017pedestrian} for details), the flow
rate was not found to vary substantially when there were fewer people
left in the room. Secondly, contrary to the assumption of uncorrelated
time gaps, we recently demonstrated that an alternation between short
time gaps and longer ones is generally expected in bottleneck flows
\cite{nicolas2017origin}. Nevertheless, these are short-time (anti)correlations
and they should not very substantially affect the results (where long
clogs play an important role). In any case, the problem can be remedied
easily, by reasoning on the distribution of time lapses for a handful of successive
escapes (\emph{e.g.}, $\sum_{p=0}^{3}\Delta t_{i+p}$), instead of considering
the individual time gap $\Delta t_{i}$.

\section{\label{sec:validity}Validity and limits of the micro-macro relation\label{sec:validation}}

In this section, we outline a validation test for the \emph{micro-macro
relation} relying on model pedestrian dynamics. More precisely, the
model under consideration is a cellular automaton that we designed
to reproduce the dynamics of competitive escape through a narrow door
observed experimentally by Zuriguel et al. \cite{Pastor2015experimental},
in particular the heavy-tailed, power-law-like distribution of time
gaps, \emph{viz.}, $p(\Delta t)\sim\Delta t^{-\alpha}$ for large
$\Delta t$, with $\alpha>0$.

\subsection{Succinct description of the model}

In the model, space is discretised into a regular rectangular grid,
in which each cell hosts at most one agent. At each time step, all
pedestrians start by selecting the site that they target among the
adjacent cells (denoted $k$ and including the present one). A move
from site $i$ to site $j$ has probability
 $p_{i\rightarrow j}\equiv e^{\frac{A_{j}-A_{i}}{x}}/\sum_{\left\langle i,k\right\rangle }e^{\frac{A_{k}-A_{i}}{x}},$
where $A_{k}$ is the attractiveness of the site (how close it is
to the exit) and the noise intensity $x=1$ has been introduced to
avoid the artefacts caused by strictly deterministic moves. If the
target site is occupied or if other people have selected it as target
site and are therefore competing for it, the agent simply waits. Otherwise,
(s)he moves to the target site. Following this round of motion, some
sites have been vacated. This allows other agents to move to their
target cell. The round is iterated until all possibilities of motion
have been exhausted. The limit of strong friction considered in the
model is noteworthy: As soon as two or more agents are competing for
a site, the conflict is sterile and nobody can move. Further details
can be found in \cite{nicolas2016statistical}.

Simulations showed that this model is able to capture the exponential
distributions of burst sizes (\emph{i.e.}, egresses in rapid succession).
But adding one last ingredient was crucial in order to replicate the
heavy tails in $p(\Delta t)$. Indeed, some amount of disorder was required. This
was achieved by introducing behaviours: Each agent ($i$) is endowed
with a propensity to cooperate $\Pi_{i}\in]0,1[$. At each time step,
this propensity determines whether agent $i$ cooperates (which occurs
with probability $\Pi_{i}$) or not. Nothing changes if the agent
behaves cooperatively. In the opposite case, the agent is impatient
to move to another site, so (s)he undervalues the attractiveness of
the current site, \emph{viz., }
$A(x_{i},y_{i}) \longrightarrow A(x_{i},y_{i})+\frac{1}{2}\ln\Pi_{i}$. 
With this additional ingredient, the model succeeded in describing
a `faster-is-slower' effect \cite{helbing2000simulating} when the crowd becomes
more impatient, but also in yielding power-law tails in $p(\Delta t)$ for impatient crowds
and narrow doors \cite{nicolas2016statistical}.

\subsection{Validation of the micro-macro relation}
For each distribution of behaviours $\Pi_i$ and each door width, a large number of evacuation realisations were simulated for an arbitrary number
of occupants $N$. The resulting histogram of total evacuation times is plotted as cyan bars in the 
right box of Fig.~\ref{fig:micro-macro}. In parallel, the distribution of time gaps
$p(\Delta t)$ was computed from the simulation of an evacuation involving a very large crowd.
On this basis, the global distribution of evacuation times predicted using the \emph{micro-macro relation} is shown
in the same figure. Clearly, there is excellent agreement between the prediction line and the histogram, which
gives credence to the validity of the \emph{micro-macro relation}.

\subsection{Limits to the validity}
However, the proposed relation was found to fail in some conditions, and sometimes severely so.
This notably happened when a process of social contagion was incorporated into the model, whereby the neighbours
of an impatient agent tended to behave more selfishly. At high contagion strengths, system-spanning correlations emerged
and the crowd sometimes ended up being extremely impatient (thus evacuating more slowly in the model), whereas it remained
in its initial, patient state during other realisations, thus evacuating faster. Such scenario splitting completely undermined the \emph{micro-macro relation}.

More generally, we also expect extrinsic fluctuations, as defined in Section~\ref{sec:origins}, to lead to a broader distribution
of global evacuation times than the micro-macro prediction. Despite these limitations, by accounting for statistical fluctuations,
the latter prediction is more informative regarding the probability of dangerously long evacuations than just the exit capacity.

\subsection{Conclusion}
In this contribution we have argued that estimates for the time spent at doors, gates or in corridors during an evacuation should
not be based exclusively on the mean flow rate at the bottleneck under study, \emph{i.e.}, its capacity. Indeed, these times may fluctuate considerably, especially
when the relative narrowness of the bottleneck and the competitiveness of the crowd favour highly intermittent dynamics. We have suggested a simple way to assess 
the extent of these fluctuations (more precisely, their statistical component) on the basis of the `microscopic' statistics of time gaps between successive escapes,
which could be collected in practice. Putting the proposed relation to the test with a cellular automaton model \cite{nicolas2016statistical} has confirmed its validity under some conditions, but also pointed to its limitations: Fluctuations are underestimated when there are strong correlations in the pedestrians' behaviours.

\emph{Acknowledgments -- }
I would like to thank Sebasti{\'a}n Bouzat and Marcelo Kuperman, who co-authored the paper \cite{nicolas2016statistical} on which this
talk was largely based.

%\bibliographystyle{spmpsci}
%\bibliography{/home/alexandre/Documents/BiblioCrowds}

\end{document}